\documentclass[usenatbib,letter]{mn2e}
\usepackage{amsmath}
\usepackage{amssymb}
\usepackage[pdftex]{graphicx}
\newcommand{\beq}{\begin{eqnarray}} 
\newcommand{\eeq}{\end{eqnarray}} 
 
\newcommand{\apj}{ApJ} 
\newcommand{\apjs}{ApJS} 
\newcommand{\apjl}{ApJL} 
\newcommand{\aj}{AJ} 
\newcommand{\mnras}{MNRAS}

\newcommand{\nat}{Nature} 
\newcommand{\na}{New Astron.}

\newcommand{\hMpc}{{\ifmmode{h^{-1}{\rm Mpc}}\else{$h^{-1}$Mpc }\fi}} 
\newcommand{\hGpc}{{\ifmmode{h^{-1}{\rm Gpc}}\else{$h^{-1}$Gpc }\fi}} 
\newcommand{\hmpc}{{\ifmmode{h^{-1}{\rm Mpc}}\else{$h^{-1}$Mpc }\fi}} 
\newcommand{\hkpc}{{\ifmmode{h^{-1}{\rm kpc}}\else{$h^{-1}$kpc }\fi}} 
\newcommand{\hMsun}{{\ifmmode{h^{-1}{\rm {M_{\odot}}}}\else{$h^{-1}{\rm{M_{\odot}}}$}\fi}} 
\newcommand{\hmsun}{{\ifmmode{h^{-1}{\rm {M_{\odot}}}}\else{$h^{-1}{\rm{M_{\odot}}}$}\fi}} 
\newcommand{\Msun}{{\ifmmode{{\rm {M_{\odot}}}}\else{${\rm{M_{\odot}}}$}\fi}} 
\newcommand{\msun}{{\ifmmode{{\rm {M_{\odot}}}}\else{${\rm{M_{\odot}}}$}\fi}} 
\newcommand{\LCDM}{$\Lambda$CDM}

\begin{document}
\title[The assembly of the Local Group]{The dark matter assembly of the
  Local Group in constrained cosmological simulations of a $\Lambda$CDM universe}
\author[J.E. Forero-Romero et al.]
{
Jaime E. Forero-Romero $^{1}$, 
Yehuda Hoffman $^{2}$,
Gustavo Yepes $^{3}$,\and Stefan Gottl\"ober $^{1}$, 
Robert Piontek $^{1}$,
Anatoly Klypin $^{4}$ 
and Matthias Steinmetz $^{1}$
\vspace*{6pt} \\ 
$^{1}$Leibniz-Institut f\"ur Astrophysik Potsdam (AIP), An der Sternwarte 16, D-14482, Potsdam, Germany\\
  $^{2}$Racah Institute of Physics, Hebrew University, Jerusalem 91904, 
  Israel\\ 
  $^{3}$Grupo de Astrof\'{\i}sica, Universidad Aut\'onoma de Madrid, 
  Madrid E-28049, Spain \\ 
  $^{4}$Department of Astronomy, New Mexico State University, Box 30001, 
  Department 4500, Las Cruces, NM 880003, USA\\ 
}
\maketitle

\begin{abstract}

We make detailed theoretical predictions for the  assembly properties
of the Local Group (LG) in the standard $\Lambda$CDM cosmological model. We
use three cosmological N-body dark matter simulations from the Constrained Local Universe
Simulations (CLUES) project, which are designed to reproduce the main
dynamical features  of the matter distribution down to the scale of a few Mpc around the
LG. Additionally, we use the results of an unconstrained 
simulation with a  sixty times larger volume to calibrate the
influence of cosmic variance.  We  characterize the Mass Aggregation
History (MAH) for each halo by three characteristic times, the
formation, assembly and last major merger times.  A major merger is
defined by a minimal mass ratio of $10:1$.    

We find that
 the three LGs share a similar MAH with
formation and last major merger epochs placed on average $\approx 10-12$
Gyr ago. Between $12\%$ and $17\%$ of the halos in the mass range
$5\times 10^{11}\hMsun <M_{h} < 5\times 10^{12}\hMsun$ have a similar MAH. In
a set of pairs of halos within the same mass range, a fraction of $1\%$ to
$3\%$ share similar formation properties as both halos in the simulated LG.
An unsolved question posed by our results is the dynamical origin of the MAH
of the LGs.  The isolation criteria commonly used to define  LG-like halos in
unconstrained simulations do not narrow down the halo population into a set
with quiet MAHs, nor does a further constraint to reside in a low
density environment.

The quiet MAH of the LGs provides a favorable environment
for the formation of disk galaxies like the Milky Way and M31. The timing for
the beginning of the last major merger in the Milky Way dark matter
halo matches with the gas rich merger origin for the thick component
in the galactic disk.  Our results support the view that the specific
large and mid scale environment around the Local Group play a critical
role in shaping its MAH and hence its baryonic structure 
at  present.

\end{abstract}

\section{Introduction}

Observations of the Milky Way (MW) and the galaxy M31 shape to a great extent our
understanding of galaxy formation and evolution. 
In particular, three landmarks have been pivotal in the development of
  theoretical studies of structure formation: a) the abundance of MW
galaxy satellites that motivated one of the strongest points of tension
with the now-standard $\Lambda$ Cold Dark Matter ($\Lambda$CDM) paradigm
of structure formation \citep{1999ApJ...522...82K,1999ApJ...524L..19M}, b) the spatial
distribution of the same satellites which triggered discussions on how
unique the host dark matter halo of the MW is 
\citep{2009MNRAS.394.2223M} and c) the measurements of
the tidal debris of disrupted merging galaxies around the MW
and M31 galaxy, confirming the  hierarchical nature of
galaxy evolution, one of the fundamental characteristics of
$\Lambda$CDM \citep{2009Natur.461...66M}. However, inferring general
conclusions on galaxy evolution based on observations of these two
galaxies requires an assessment on how biased the properties of the MW and
M31 are with respect to a given control population.

In the framework of $\Lambda$CDM, the study of the MW and M31 starts by
modeling their individual host dark matter halos, \emph{assuming} that their simulated
formation histories are "typical", or at least compatible with the assembly of the
real Local Group \citep{2009MNRAS.395..210D,2010MNRAS.406..896B}.  The basic definition of a
LG (in terms of the dark matter distribution) has two basic elements based on
the state of the system \emph{today}: (i)  the estimated masses of the dark matter
halos corresponding to the MW and M31  (see for instance 
  \cite{2010MNRAS.406..264W} and references therein) and (ii) the isolation of these
two halos from other massive structures
\citep{2004AJ....127.2031K}. Two additional constraints could be the
separation and the relative velocity of the two halos \citep{2005ApJ...635L..37R}. However, the condition
on the LG isolation admits a strict formulation, by requiring that the
environment, in terms of the mass and position of the dominant galaxy
clusters in the Local Universe, be as close as possible to the one
inferred from observations. Such an additional condition imposes
restrictions on the possible outcomes of structure formation on scales
of the order of $\sim5$ Mpc. This is considered here as the meso-scale
as opposed to the large ($\gtrsim 5$ Mpc) or the small ($\lesssim 1$
Mpc) scales.

The new feature in the analysis presented in this paper is the
inclusion of such observational constraints around the LG environment
in the initial conditions of the simulation. In a series of three
simulations from such initial conditions, in a WMAP5 cosmology with a
normalization $\sigma_{8}=0.817$ \citep{2009ApJS..180..330K}, we are able to define a sample of
three LG dark matter halo pairs that form and evolve under specific
conditions reflecting  structure of the Local Universe. In addition we
will take advantage of one of the largest cosmological simulations
carried out to date, the Bolshoi Simulation
\citep{2010arXiv1002.3660K}, to explore a larger sample of halos within the
mass range of the LG, and calibrate possible cosmic variance effects.

We analyze the constrained simulations with the primary goal of
quantifying the assembly histories of the LG halos. This is driven by
two different motivations. One is to find out whether the simulated
LGs, that are selected by dynamical considerations pertaining to their
redshift zero structure, have mass aggregation histories (MAHs) that
lead to the formation of disk galaxies like the MW and M31.  The other
is to find out whether such a MAH is dictated by by meso-scale
environment of the LG, or whether a random  selection of objects
similar to the LG is likely to have a similar MAH.

In Section \ref{sec:clues}, we describe
our simulations and the method to re-construct the mass aggregation
histories. In Section \ref{sec:sample} we describe how we build the different
control samples for our statistical analysis. In Section \ref{sec:results} we
study the MAHs in the different samples and argue that the selection
by different isolation criteria does not induce a strong bias in the
statistics describing the MAHs. In Section \ref{sec:discussion} we discuss the
possible origin of these findings and comment on the connection with
observations of the MW and M31. In Section \ref{sec:conclusions} we summarize
our conclusions.

\section{The simulations and Mass Aggregation Histories}
\label{sec:clues}

In this paper we make use of four cosmological N-body dark matter simulations. Three
of them are part of the Constrained Local Universe Simulations
(CLUES) project \footnote{\tt http://www.clues-project.org/},
whose aim is to perform N-body cosmological simulations that reproduce the
local large scale structure in the Universe as accurately as current
observations allow. The fourth simulation is the Bolshoi Simulation,
which was performed from unconstrained initial conditions and spans a
volume  $\sim 60$ times larger than each one of the CLUES
simulations. In this section we will describe these simulations and
the procedure we have used to construct the mass aggregation histories
for the dark matter halos.

\subsection{The CLUES simulations}

First we describe the procedure employed to generate the constrained initial conditions.  The observational constraints are the peculiar velocities drawn from the
MARK III \cite{1997ApJS..109..333W}, surface brightness fluctuation
\cite{2001ApJ...546..681T}  and the position and virial properties of
nearby X-ray selected clusters of galaxies \cite{2002ApJ...567..716R}.
The 
\cite{1991ApJ...380L...5H} algorithm is used to generate
the initial conditions as constrained realizations of Gaussian random
fields.  These observational data sets impose constraints on the
outcome of structure formation on scales larger than a few
megaparsec. 

These constraints affect only the large and meso-scales of the  initial
conditions of the simulations, leaving the small scales essentially
random. In particular. the presence of a local group with two dark matter
halos roughly matching the masses, separation and relative velocities
of the MW and M31 cannot be constrained.  The strategy employed here is to 
construct an ensemble of 200 different realizations of the constrained
initial conditions and simulate these with  $256^3$ particles on a box with side length
$64$\hMpc using the Tree-PM MPI N-body code Gadget2 \citep{2005MNRAS.364.1105S}, and then scan these  for 
appropriate LG-like objects within a search box centered on the actual
position of the LG. Only three realizations are found to have such a LG object
  following the criteria detailed at the end of Sect.  \ref{sec:sample}. It follows that the
simulations analyzed here obey two kinds o selection rules. By
construction these are constrained simulations whose large and
meso-scales are designed to mimic the local Universe. Then, post
factum, the simulations that have the appropriate LGs are selected for
further analysis.

The selected simulations are then re-simulated at high resolution of $1024^3$
particles. The high resolution extension of the low-resolution simulation is
obtained by creating an unconstrained realization at the desired resolution,
fast Fourier transforming it to $k$-space and substituting the unconstrained
low $k$ modes with the constrained ones. The  resulting realization is made of
unconstrained high $k$ modes and constrained low $k$ ones. The
transitional scale happens around the length scale corresponding to
the Nyquist frequency of the $256^3$ mesh, $\lambda_{\mathrm{Ny}}=2\times
64/256$ \hMpc $=0.5 $\hMpc. This corresponds to a mass scale of
$M_{\mathrm{Ny}}\approx 1.2\times 10^{9}$\hMsun, below which the
structure formation can be considered as emerging primarily from the
unconstrained $k$ modes.

The cosmological parameters in these high resolution simulations are consistent with a WMAP5
cosmology with a  density $\Omega_{m}=0.28$, a cosmological constant
$\Omega_{\Lambda} = 0.72$, a dimensionless Hubble parameter $h=0.73$, a
spectral index of primordial density perturbations $n=0.96$ and a
normalization $\sigma_{8}=0.817$ \citep{2009ApJS..180..330K}. With these characteristics each particle has a
mass $m_{p}=1.89\times 10^{7}$ \hMsun.

\subsection{The Bolshoi Simulation}
We have used as well the Bolshoi simulation \citep{2010arXiv1002.3660K} to
verify that the constrained simulation did not bias the halo samples and their
MAHs\footnote{Halo catalogs for these simulations are available at \tt http://www.multidark.org/MultiDark/}. The simulation was done in a cubic volume  of $250$ \hMpc on a side
using $2048^3$ particles, leading to a particle mass of $m_p=1.35\times
10^{8}$ \hMsun, roughly 10 times lower than the resolution in the CLUES
simulations.

We take from the Bolshoi simulation eight non-overlapping sub-volumes. Each sub-volume has a
cubic size of  $100$ \hMpc on a side, corresponding to a comoving
volume comparable to the three CLUES simulations
combined. The halo samples in the sub-volumes will be used to
calibrate the impact of cosmic variance on the different statistics we
use to characterize the halo populations.

\subsection{Halo identification and merger tree construction}

In order to identify halos we use a FOF algorithm. We do not include
any information of the substructure in each  halo. All the
analysis related to the mass aggregation history is done in terms of
the  host halos. In particular the mergers do not correspond to the fusion
of an accreted sub-halo with a central dominant host halo, but instead
 correspond to the moment of two halos overlapping for the first time. 

The FOF algorithm has a linking length of $b=0.17$ times the mean inter
particle separation. The mean overdensity of objects found with
  this linking length at redshift $z=0$ is 680
  \citep{2011arXiv1103.0005M}. We identify the halos for $80$
  snapshots more or less equally spaced 
  over the 13 Gyrs between redshifts
  $0<z<7$. All the objects with 20 or more particles are kept in the
  halo catalogue and considered in the merger tree construction. This
  corresponds to a minimum halo mass of $M_{min}=3.78\times
  10^{8}$\hMsun. Within the CLUES simulations a Milky Way like dark
  matter halo of mass  $\sim 1.0\times 10^{12}$\hMsun\ is resolved
  with $\sim 5 \times 10^{4}$ particles, in the Bolshoi simulation it
  is resolved with  $\sim 7 \times 10^{3}$ particles.  For the Bolshoi
  simulation 
  we have used snapshots spaced by roughly $400$Myr and
  followed the exact same procedure to build the halo catalogues and the
  merger trees.

  Within the FOF analysis all FOF groups with 20 or more
  particles are identified. The merger tree construction is based on
  the comparison of the particles in FOF groups in
  two consecutive snapshots. Starting at $z=0$ for every FOF group in the
  catalog, $G_{0}$, we find all the FOF groups in the previous snapshots that
  share at least thirteen particles with $G_{0}$ and label them as tentative 
  progenitors.  Then, for each tentative progenitor, we find all the
  descendants
  sharing at least thirteen particles. Since the smallest FOF groups
  contain 20 particles at least 2/3 of the particles must be
  identified in tentative progenitors or descendants.  Only the
  tentative progenitors that have as a
  main descendant the group $G_{0}$ are labeled as confirmed
  progenitors at that
  level. We iterate this procedure for each confirmed progenitor,
  until the last
  available snapshot at high redshift. By construction, each halo in the tree
  can have only one descendant, but many progenitors. 

  The mergers of FOF groups correspond to the time where the FOF
  radii of two halos overlap for the first time. The infall of the
  less massive halo into the host and the subsequent inspiral,
  disruption and fusion will be delayed with respect to the time of
  the FOF merger. Different theoretical approximations and
  methodologies can predict the infall-fusion time-scale only as an
  order-of-magnitude estimate \citep{2010ApJ...724..915H}. The most used time-scale for this
  process is based on the Chandrasekhar dynamical friction formula,
  but improved estimates based on
  numerical simulations 
  \citep{2008MNRAS.383...93B,2010ApJ...724..915H} yield
\begin{equation}
t_{\mathrm{infall}} = 0.56\left( \frac{R_{\rm vir}}{V_{\rm vir}}\right)
\frac{(M_{\rm vir}/M_{\rm sat})^{1.3}}{\ln(1+M_{\rm vir}/M_{\rm sat})},
\end{equation}
where $R_{\rm vir}$, $V_{\rm vir}$ and $M_{\rm vir}$ are the virial radius,
velocity and mass of the host halo, $M_{\rm sat}$ the mass of the
future satellite at the moment of infall at $R_{\rm vir}$.  A median initial circularity of
the satellite orbit of $0.5$ has been assumed. For mass ratios of
$M_{\rm vir}/M_{\rm sat}=10$ 

\begin{equation}
\label{eq:infall}
t_{\mathrm{infall}} = 4.85 \left(\frac{R_{\rm vir}}{V_{\rm vir}}\right).
\end{equation}

\begin{figure*}
\begin{center}
\includegraphics[scale=0.50]{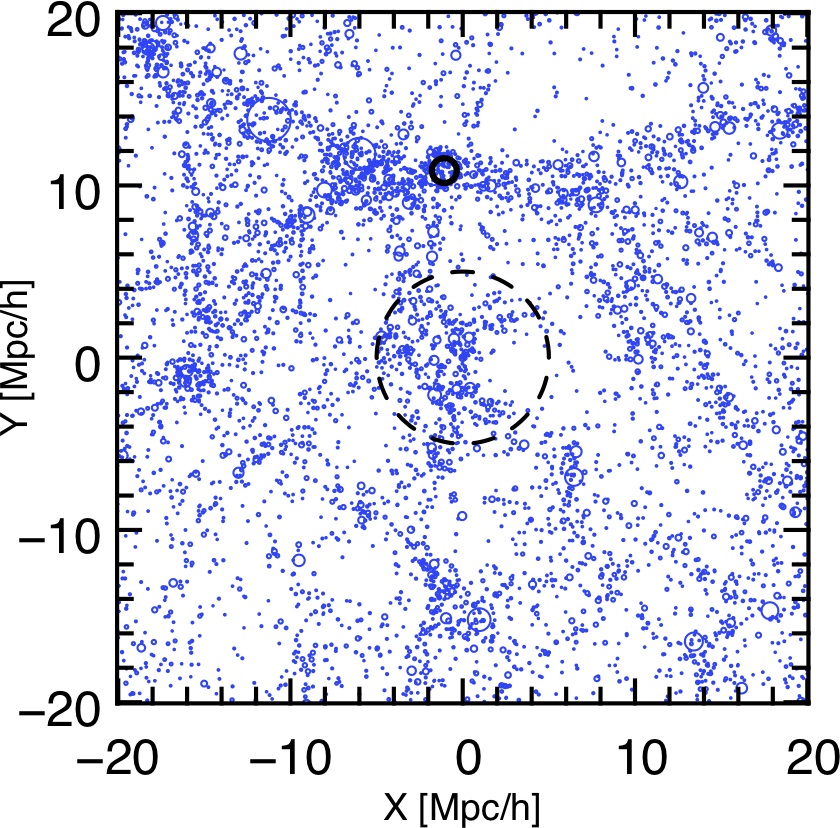}\hspace{0.5cm}
\includegraphics[scale=0.50]{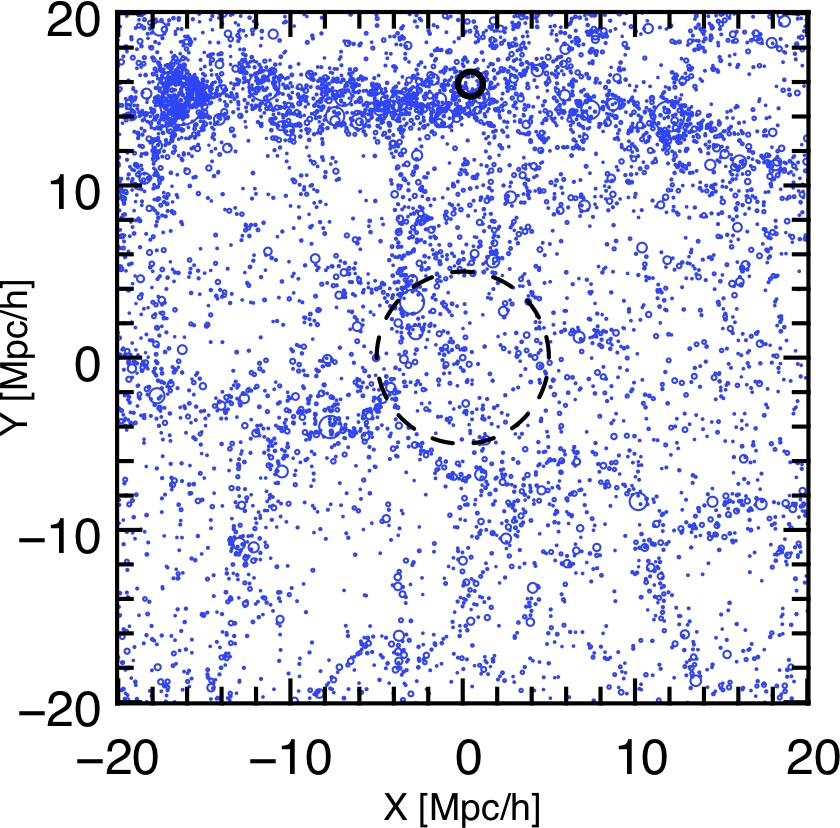}\vspace{0.8cm}
\includegraphics[scale=0.50]{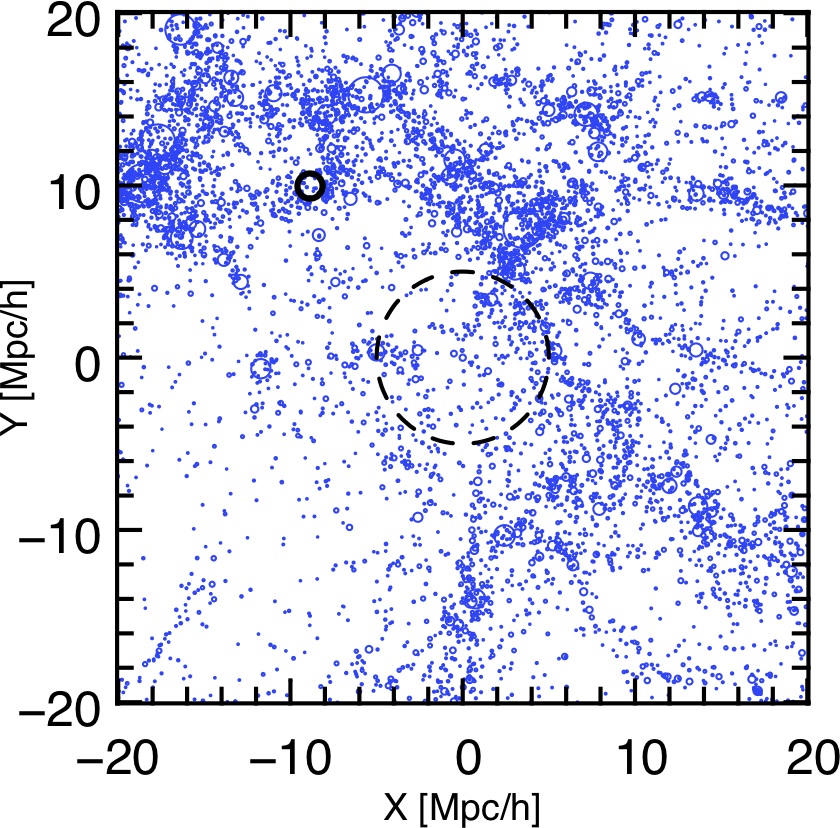}\hspace{0.5cm}
\includegraphics[scale=0.50]{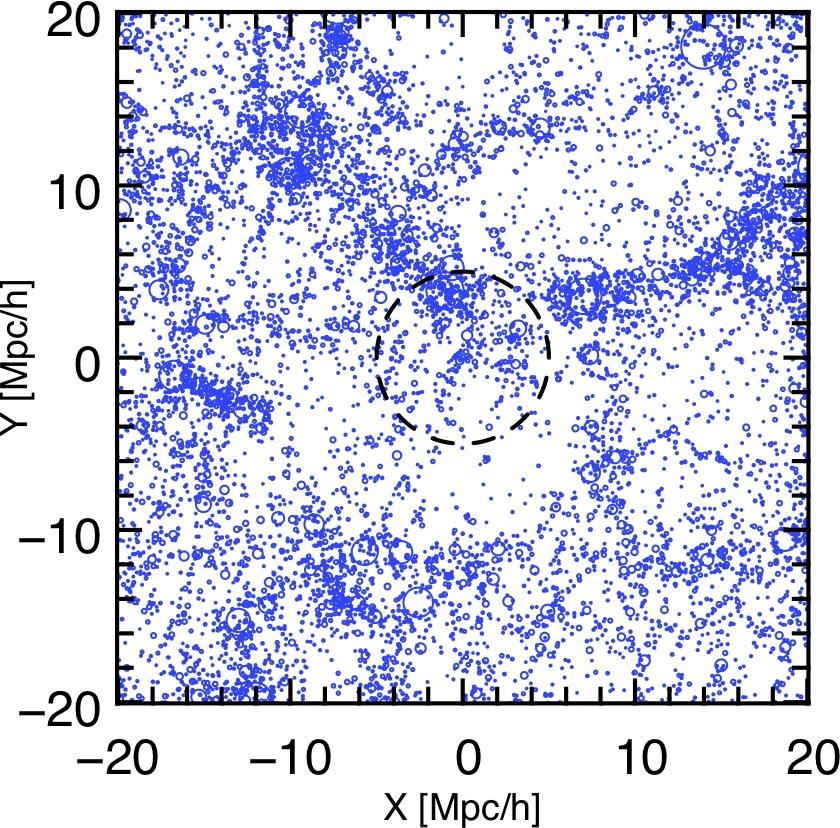}
\end{center}
\caption{\label{fig:lss}Halo distribution in the three CLUES and
  the Bolshoi (lower right)
  simulations. Only halos more massive than $M_h>
  2\times 10^{10}$\hMsun have been included. The radius of each circle corresponds to the radius
  defined by the FOF algorithm which is calculated to be the radius of a
  sphere with an equivalent volume as the FOF group. The dashed
  circle marks a $5$\hMpc environment centered in the most massive halo of the
  LG. The solid thick circle
   shows the projected position of the halo identified with the Virgo
  cluster. The cut is  $25$\hMpc thick and is centered at the LG
  position.}
\end{figure*}

\begin{table*}
\caption{\label{table:assembly} Properties of the MW-M31 pairs. Column 1:
 parent simulation. Column 2: Halo name (either MW or M31); column 3: FOF mass; column 4: last major merger time ; column 5:
 formation time; column 6: assembly time; column 7: matter
 overdensity calculated with in a sphere of $5$\hMpc. All times are look-back-times.}
\begin{tabular}{ccccccc}\hline
Simulation & Halo Name & FOF Mass & $\tau_{M}$& $\tau_{F}$ & $\tau_{A}$ & $\delta_5 + 1$\\
  & & [$10^{12}\hMsun$] &  [Gyr] & [Gyr] & [Gyr] & \\\hline
CLUES-1 & M31 & 1.39 &  11.0 & 11.0 & 11.5 & 0.72\\
CLUES-1 & MW & 0.99 &  10.0& 9.3& 9.7& 0.69\\
CLUES-2 & M31 & 0.98 & 12.0& 10.0& 10.4& 0.78\\
CLUES-2 & MW & 0.77 & 11.3& 11.0& 11.0& 0.87\\
CLUES-3 & M31 & 1.45 & 11.0& 10.6& 11.0& 0.75\\
CLUES-3 & MW & 1.11  & 9.8& 9.8& 11.0& 0.80\\\hline
Average &  & 1.15 &10.9 & 10.3 & 10.8 & 0.76\\
Standard Deviation &  & 0.23 &0.8 & 0.6& 0.6& 0.05\\\hline
\end{tabular}
\end{table*}

\subsection{Local Group selection}
A LG in a constrained simulation consists of two 
  main halos within a certain mass range, within a distance range
  and obeying some isolation conditions\footnote{A quantitative description of
  these conditions is presented at the end of Section \ref{sec:sample}.}. In addition it should reside 
  close to the relative position of the LG with respect to the Virgo cluster. Given the periodic boundary conditions of the
  simulations and the lack of treatment of the Zeldovich linear displacement in
  the reconstruction of the initial conditions, the large scale structure of the
  simulations is displaced by a few Megaparsecs among different
    realizations of the simulation. The most robust features
  of the constrained simulations are the Virgo cluster and the  Local
  Supercluster.  Their positions in the initial conditions are known,
at $z=0$ their environment is searched for halos in the corresponding mass
range to determine their present positions.
These are used to fix the  'position' of the simulation in
  relation to the actual universe. In Table
  \ref{table:assembly} we summarize the masses of the MW and M31 halos
  identified by the FOF halo finder in these three simulations.

Figure \ref{fig:lss} shows the large scale structure of the three
constrained realizations centered on the position of the LG in
each box in a slice 25\hMpc thick.  In the three
CLUES simulations shown in Fig.  \ref{fig:lss} the projected position
of the Virgo cluster is shown by a thick circle. The fourth panel in
the same figure shows a cut of the same geometrical characteristics
from the Bolshoi simulation, centered on one LG-like object.

\subsection{Merger trees description}

For each  merger tree we define three different times to characterize
the MAHs. Each time has direct connection with the expected properties
of the baryonic component in the halo. The times, measured as
look-back time in Gyr, are:   

\begin{itemize}
\item{{\bf Last major merger time} ($\tau_M$):  defined as the time
    when the last FOF halo interaction with ratio 1:10 starts. This
    limit is considered to be the mass ratio below which the
    merger contribution to the bulges can be estimated to  be $< 5\%$-$10\%$
    \citep{2010ApJ...715..202H}. Strictly speaking, as we do not
    follow     sub-structure in the simulation, this event corresponds
    to the time when the merger fell into the larger halo and for
    the first time     became a sub-halo. One can use
    Eq. (\ref{eq:infall}) to estimate the infall time-scale of the
    satellite to the center of the host.}  
\item{{\bf Formation time}  ($\tau_F$): marks the time when the main branch in the tree
    reached half of the halo mass at $z=0$. This marks the epoch when
    approximately half of the
    total baryonic content in the halo could be already in place in a virialized
    object.}
\item{{\bf Assembly time} ($\tau_A$): defined as the time when the
    mass in progenitors more massive than M$_f=10^{10}$\hMsun is half
    of the halo mass at $z=0$. This time is related to the epoch of
    stellar component assembly, as    the total stellar mass depends
    on the integrated history of     all progenitors
    \citep{2006MNRAS.372..933N,2008MNRAS.389.1419L}. The exact
    value of $\tau_{A}$ is dependent on $M_f$, the specific value selected in this
    work was chosen to allow the comparison of assembly times against
    the results of the Bolshoi simulation which has a lower mass
    resolution than the CLUES volumes.} 

\end{itemize}

\begin{figure}
\begin{center}
\includegraphics[scale=0.50]{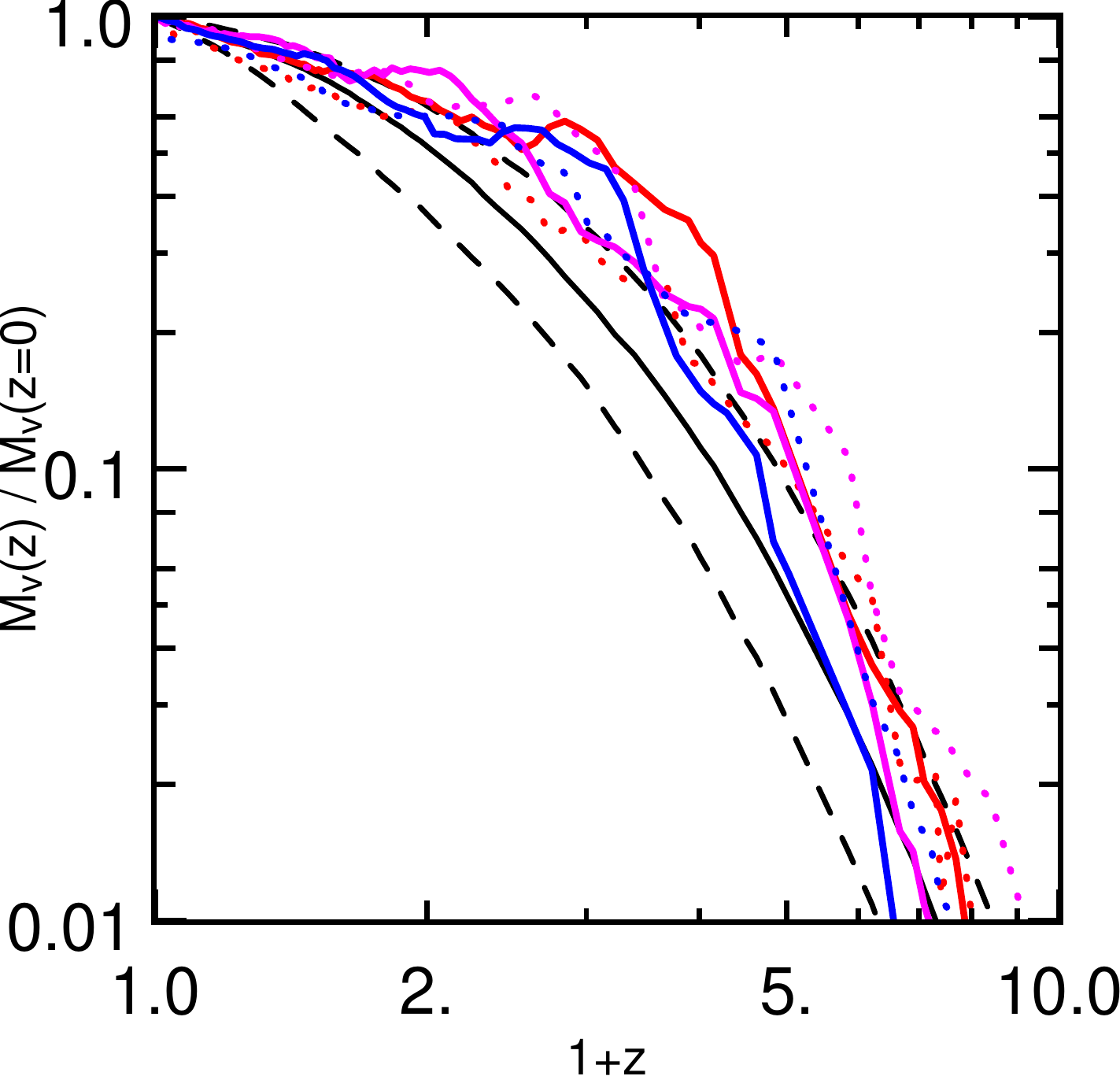}
\end{center}
\caption{\label{fig:mah}  Mass assembly histories of LG halos in the CLUES
  simulation as a function of redshift. The solid black line shows the median
  MAH for all halos in the CLUES simulations within the mass range $5.0\times
  10^{11}\hMsun< M_{h}<5.0\times 10^{12}\hMsun$, the dashed lines show  the
  first and third quartiles. Also plotted as colour lines are the MAHs for the
  MW (dotted) and M31 halos (continuous)  in the three constrained
  simulations. The assembly history for the LG halos is systematically
  located over the median values as sign of early assembly with
  respect to all halos in the same mass range.}
\end{figure}

In Table \ref{table:assembly} we summarize the values of these three
  different times for the three pairs of MW-M31 halos.  In Figure
  \ref{fig:mah} we show the median mass aggregation history in the
  main branch as a function of redshift for halos in the mass range
  $5.0\times 10^{11}\hMsun< M_{h}<5.0\times 10^{12}\hMsun$.  Following
  \cite{2010MNRAS.406..896B} we fit the MAH by a function of the kind 
\begin{equation}
M(z) = M_{0}(1+z)^{\beta}\exp(-\alpha(\sqrt{1+z} -1 )),
\end{equation}
with $\alpha=4.5$ and $\beta=2.24$. These values provide a good fit within
$2.3\%$ for $z<7$. In the same Figure we overplot the main branch
growth for the six halos in the three simulated LGs. The MAHs of
these halos are systematically located above the mean, an indicator of
early matter assembly with respect to the halos within the same mass
range.

\section{Selection of  Local Groups and Control  Samples}
\label{sec:sample}

\begin{table*}
\caption{\label{table:samples} Names and description of the four samples used
 to quantify the formation history of the LG halos. The three first samples 
 are constructed both from the CLUES and Bolshoi simulations. By definition
 the {\it LG} sample can only be constructed from the CLUES simulations.
 The size refers to the total number of objects in the corresponding volume (individual
 halos or pairs).}
\begin{tabular}{llcc}\hline
Name & Description & Size (CLUES)& Size (Bolshoi)\\\hline
{\it Individuals} & All the distinct halos in the mass range $5.0\times
10^{11}$\hMsun - $5.0\times 10^{12}$\hMsun & 4278 & 88756\\
{\it Pairs} & All the pairs of halos constructed from the {\it
  Individuals} sample. & 1101 & 21877 \\
{\it Isolated Pairs}& Subset from the {\it Pairs} sample
following some isolation criteria (see \S\ref{sec:sample})& 85 & 1785\\
{\it LG}& The three pairs of LG halos from the constrained simulations. & 3&---\\\hline
\end{tabular}
\end{table*}

\begin{figure*}
\begin{center}
\includegraphics[scale=0.33]{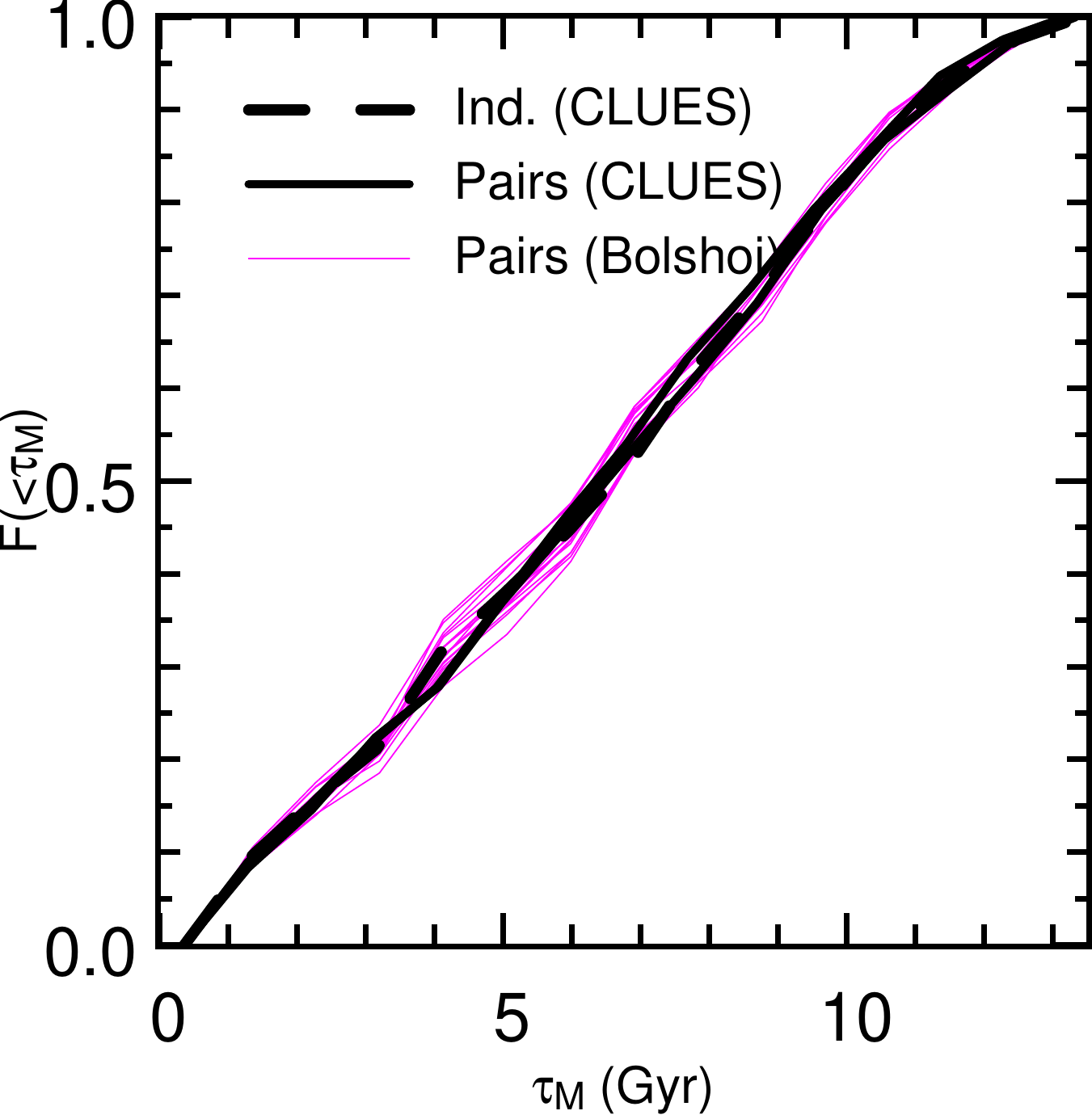}\hspace{0.5cm}
\includegraphics[scale=0.33]{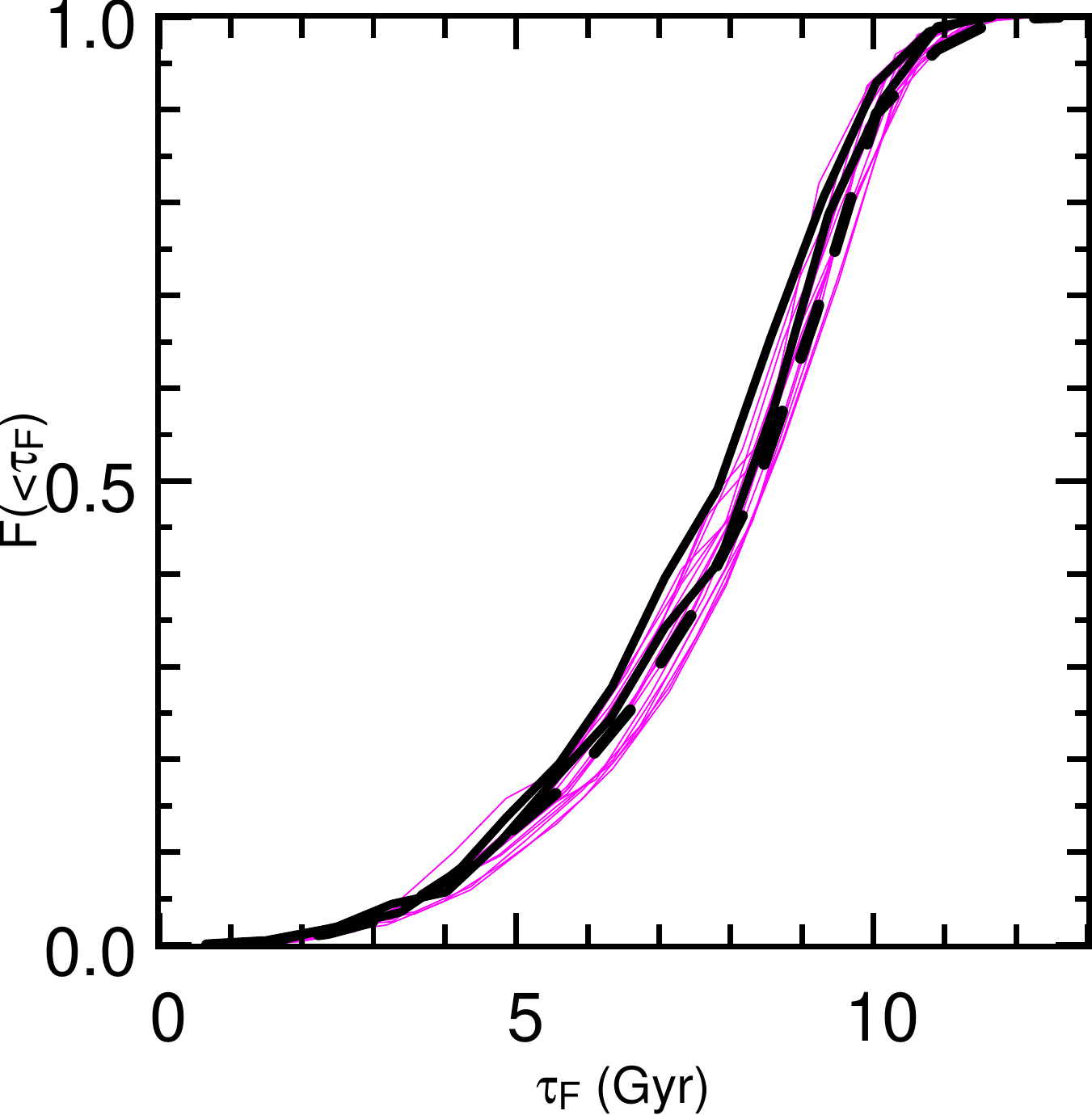}\hspace{0.5cm}
\includegraphics[scale=0.33]{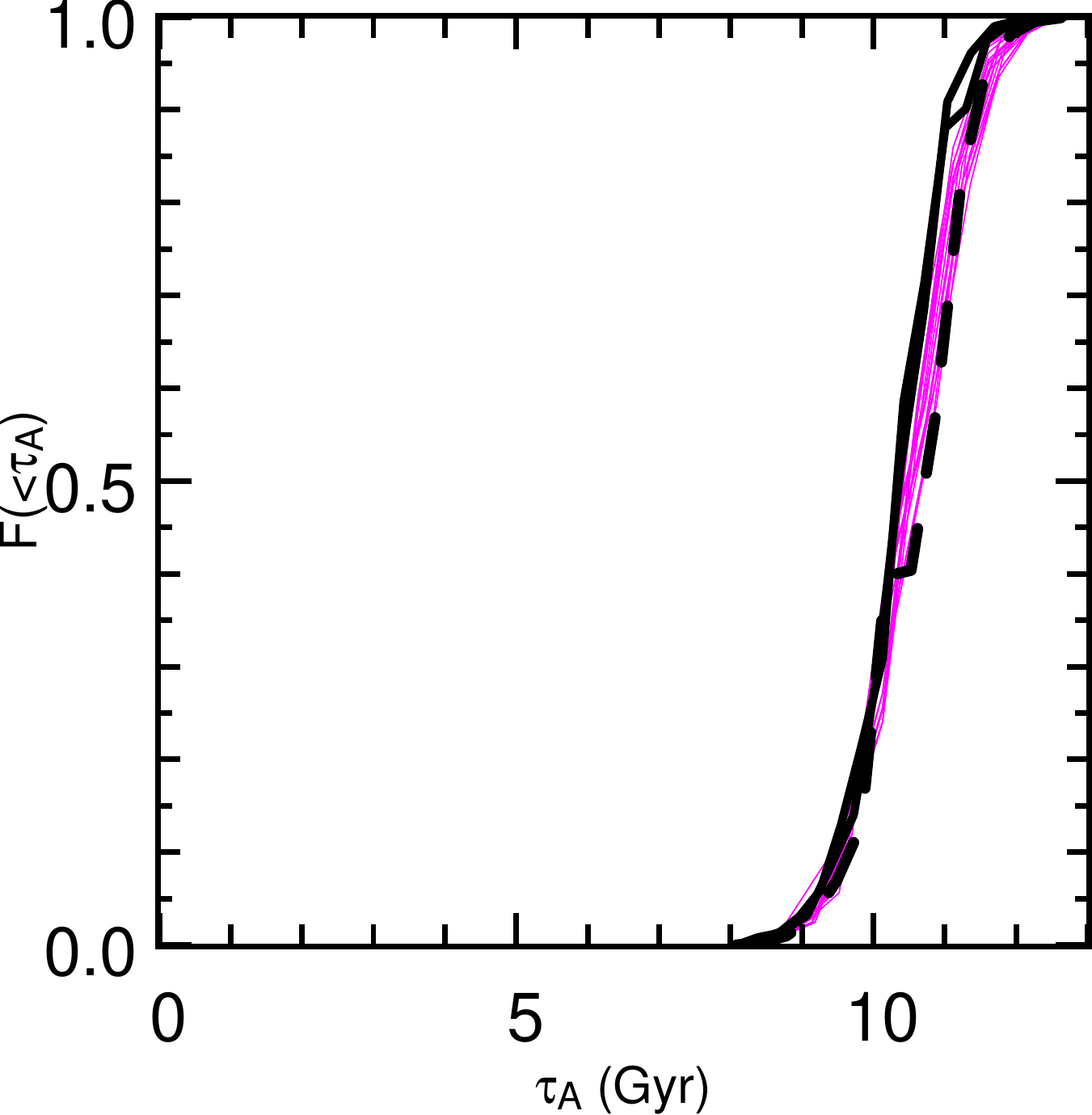}
\end{center}
\caption{\label{fig:integrated_A_B} Fraction of halos with merger histories
  described by a MAHs with $\tau_M$, $\tau_F$ and $\tau_A$ larger than a given
  value. The lines represent different samples. The sample of {\it
  Individuals} (dashed) and {\it Pairs} (thick continuous lines) from the CLUES
  simulations and the {\it Pairs} extracted from eight sub-volumes in the
  Bolshoi simulation (thin continuous lines).}
\end{figure*}

\begin{figure*}
\begin{center}
  \includegraphics[scale=0.45]{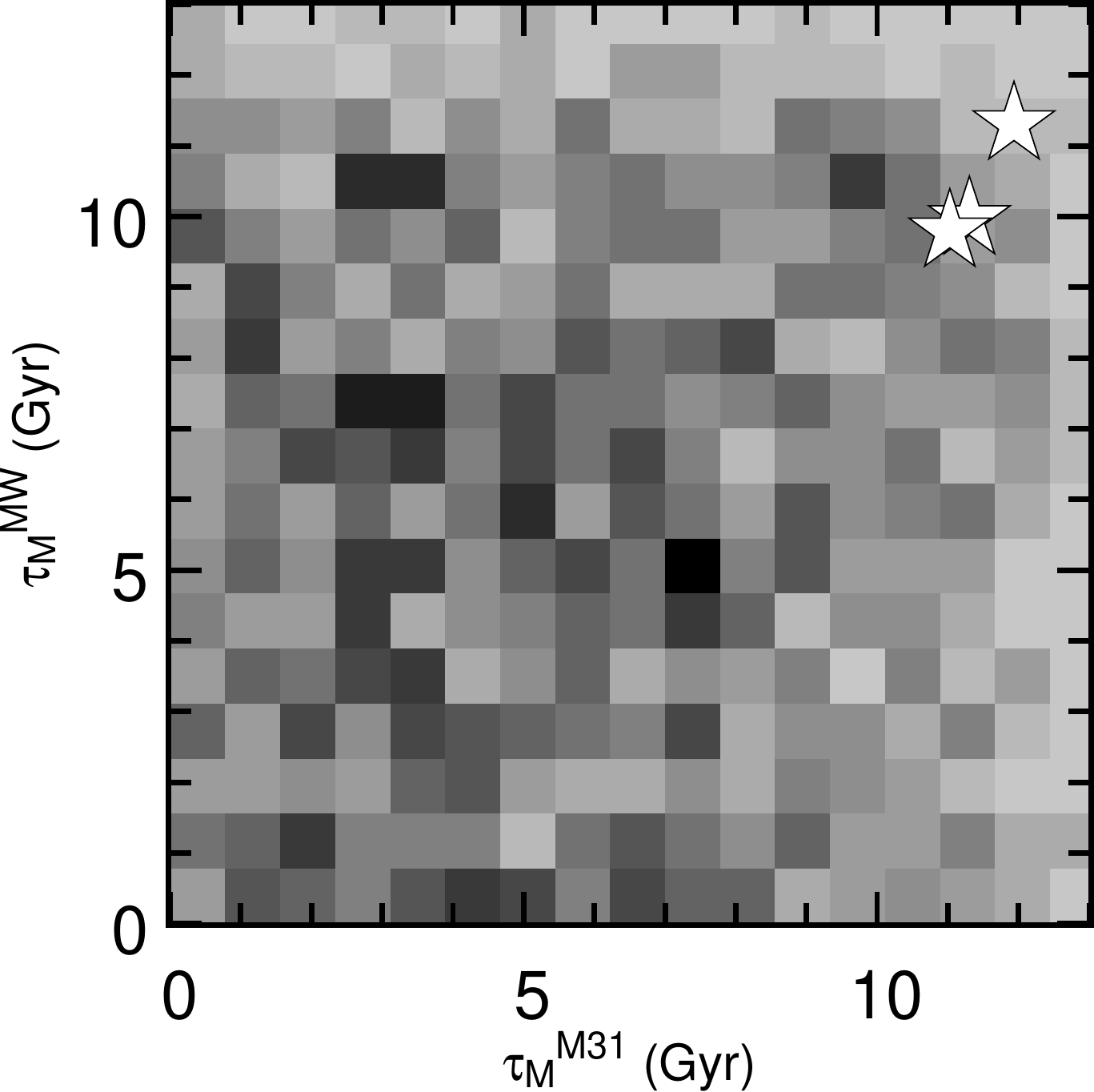}\hspace{0.5cm}
\includegraphics[scale=0.45]{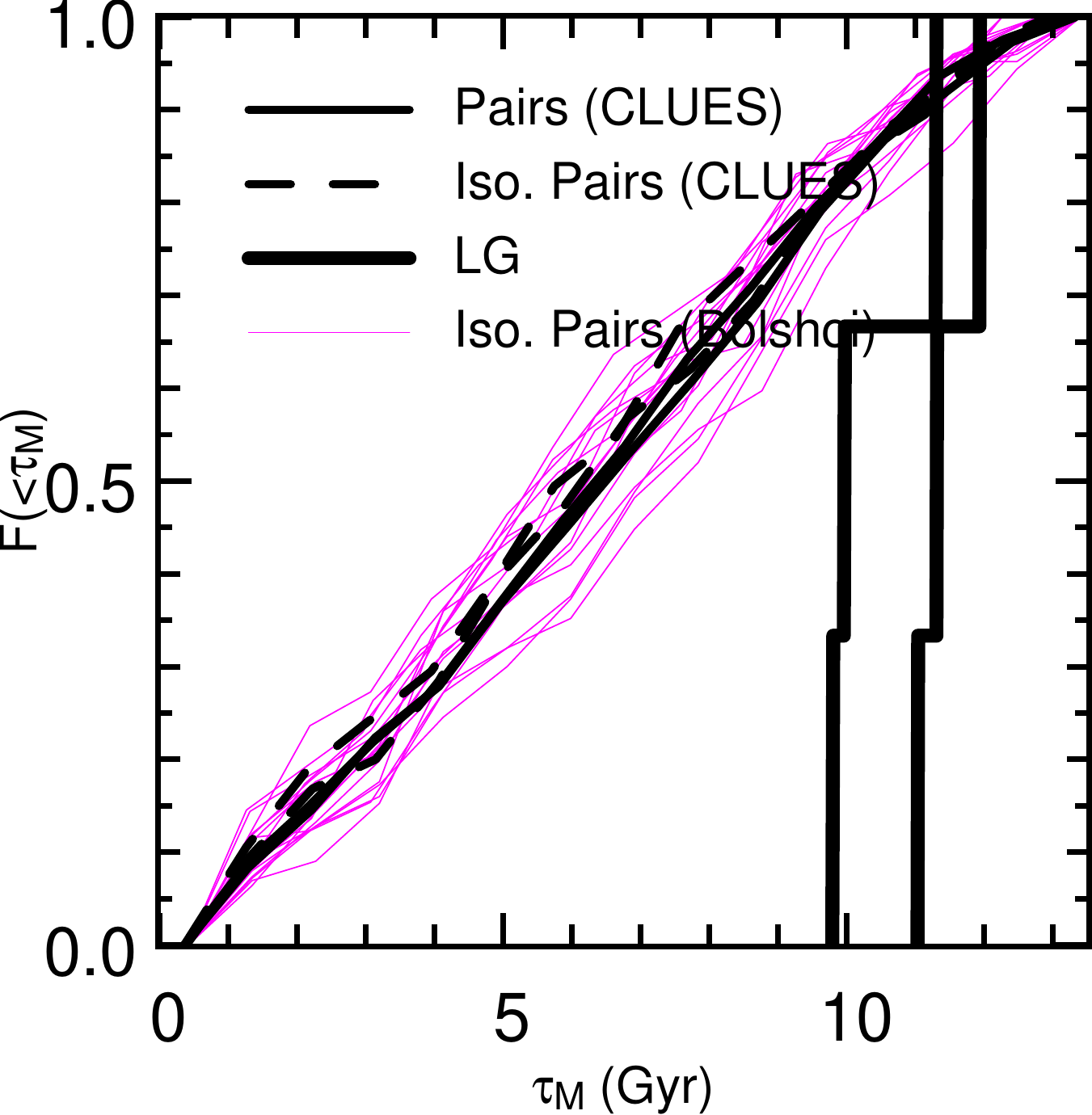}
  \includegraphics[scale=0.45]{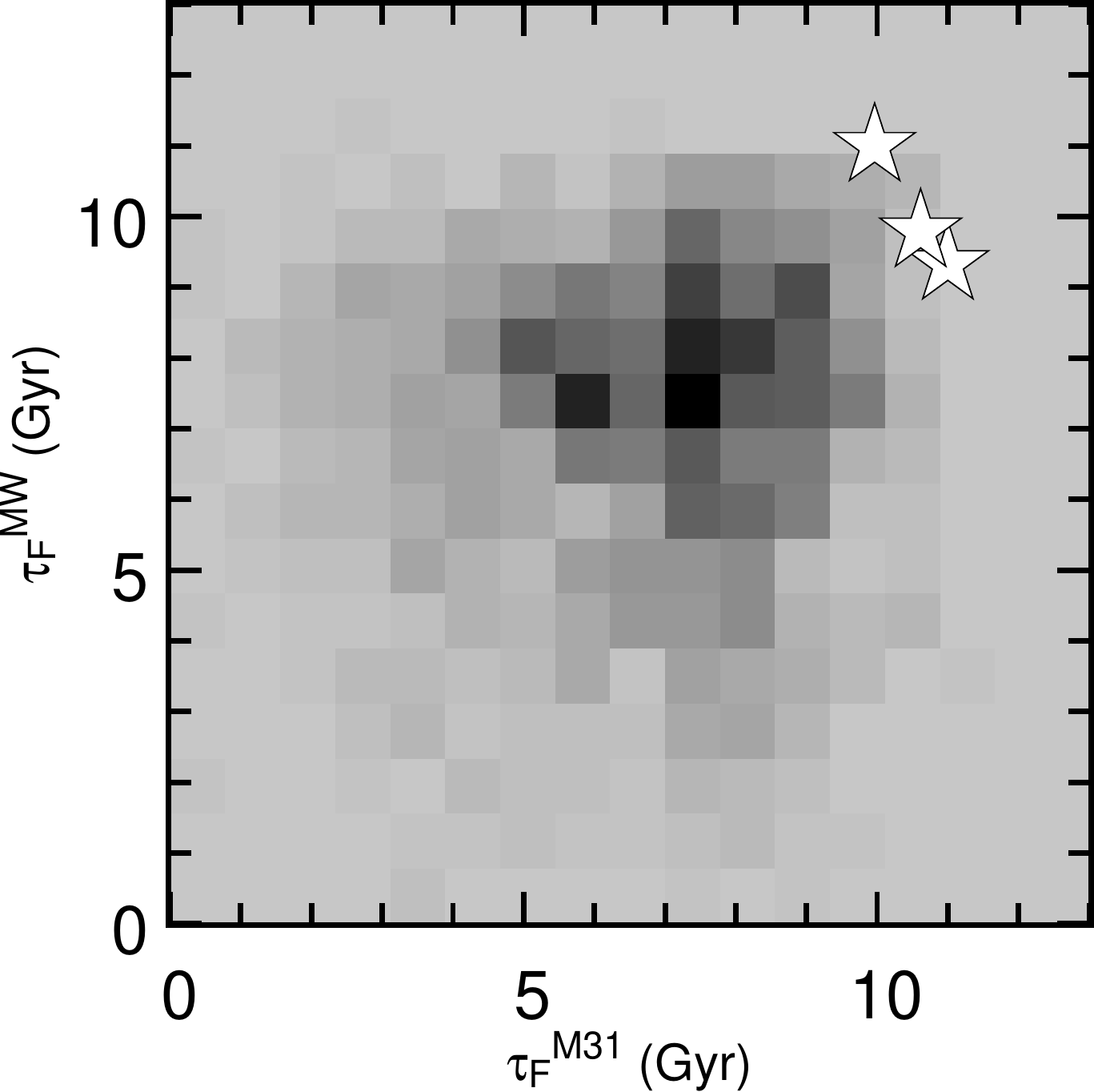}\hspace{0.5cm}
\includegraphics[scale=0.45]{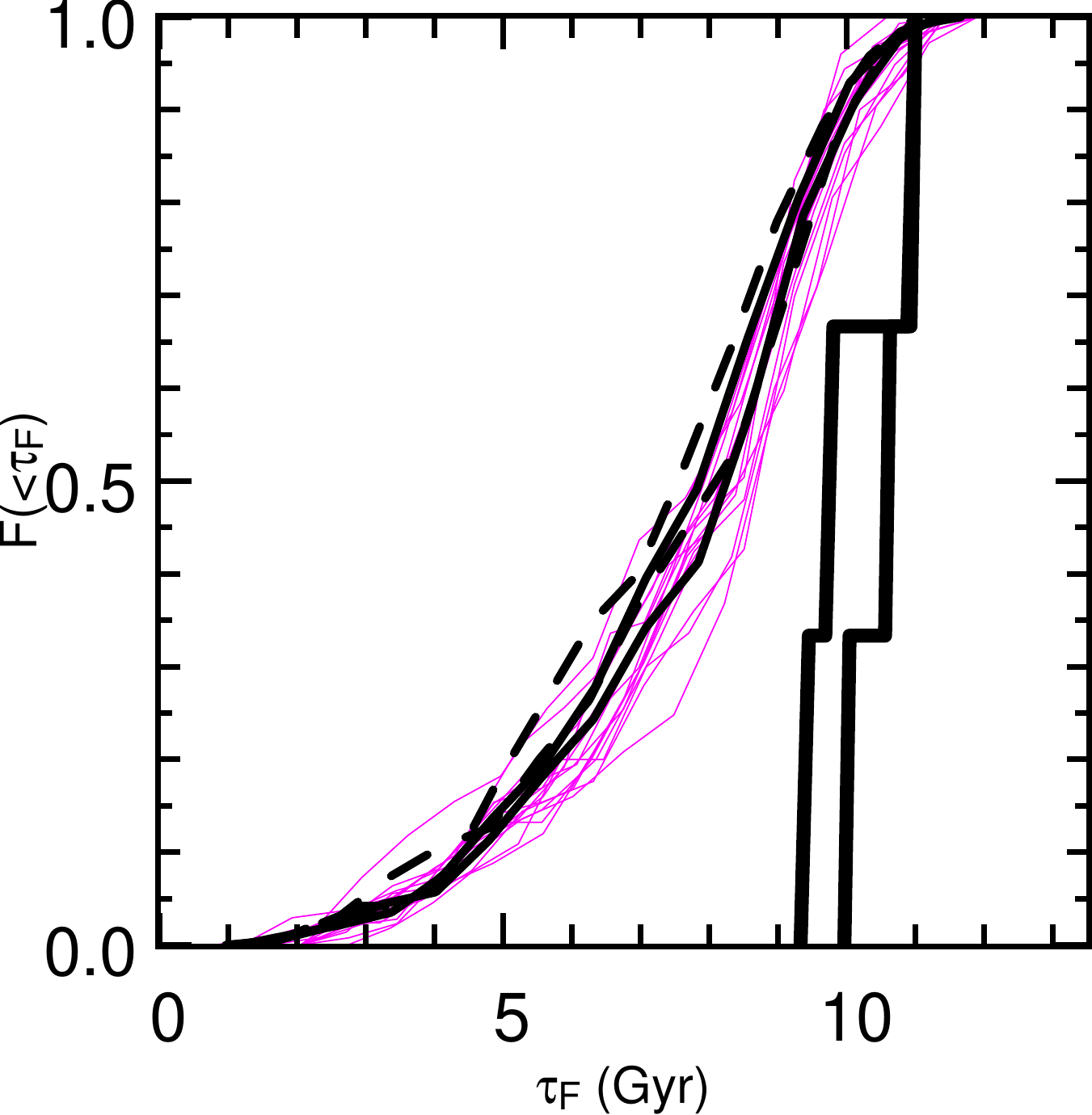}
\includegraphics[scale=0.45]{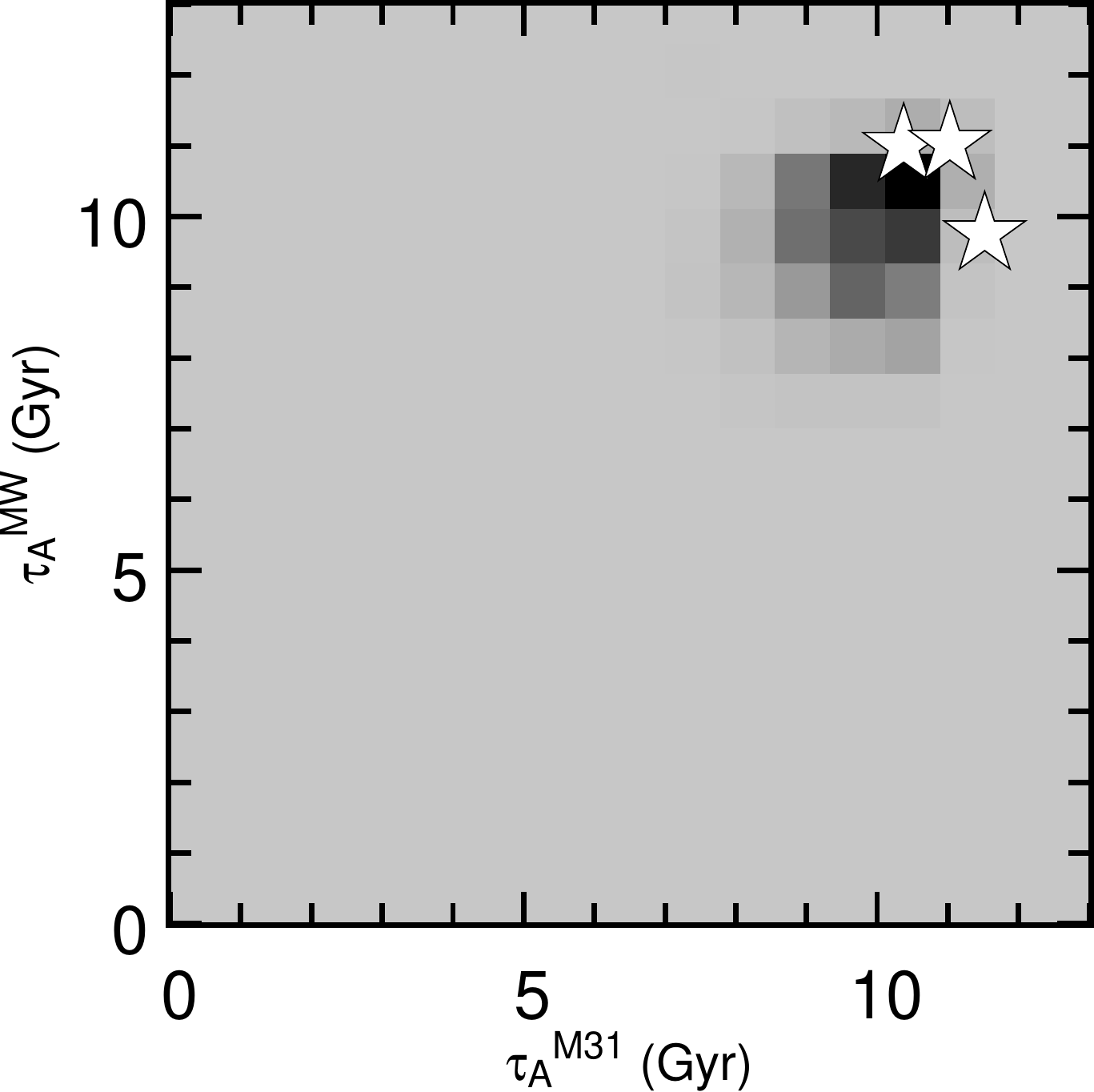}  \hspace{0.5cm}
\includegraphics[scale=0.45]{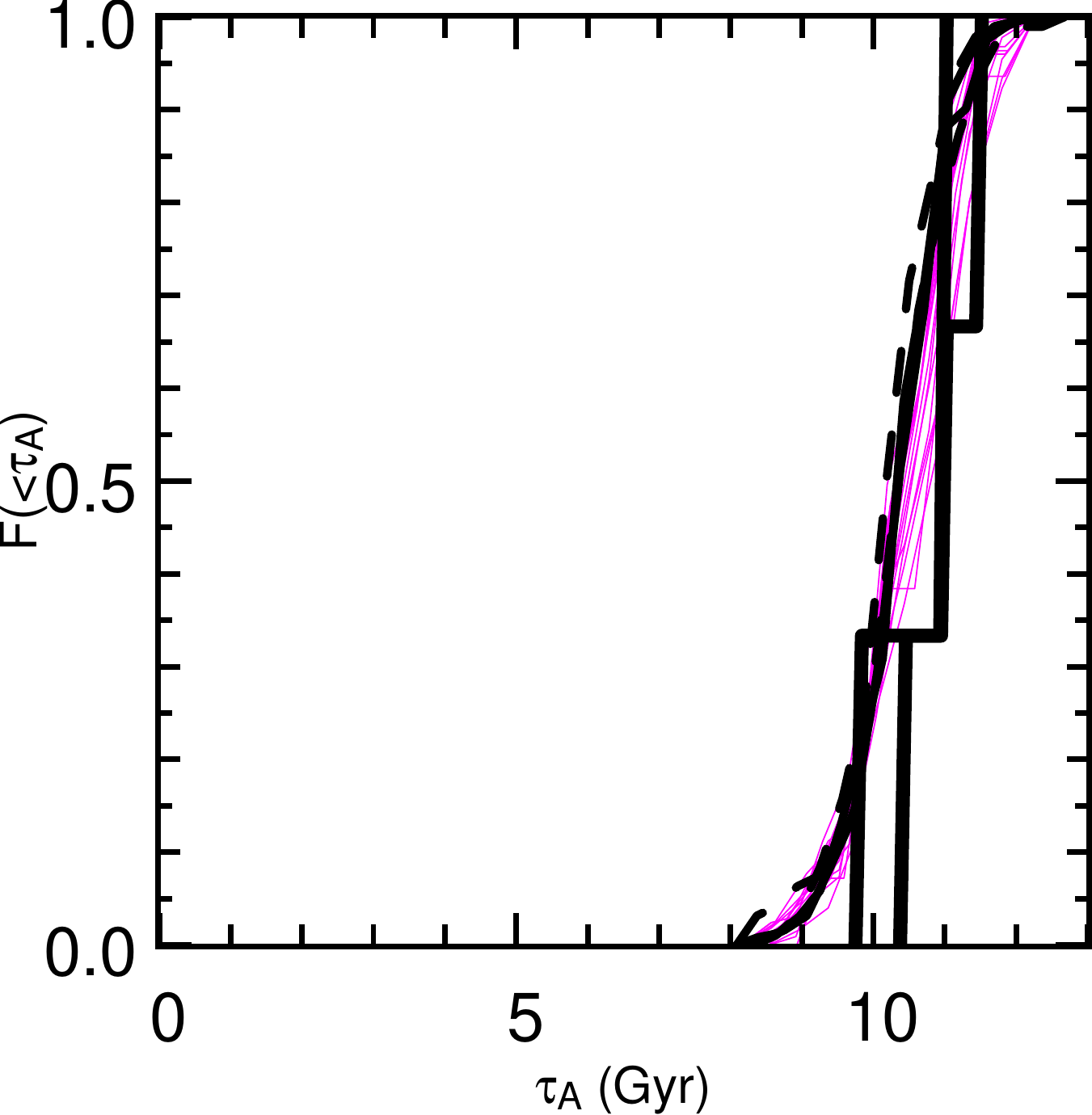}
\end{center}
\caption{\emph{Left column}. Joint distributions of three different times
  (last major merger, formation and assembly) describing the mass aggregation histories.
 Each point in the plane  represents a pair MW-M31 with histories described by the time values at that
 point. Levels in shading coding indicate the number of halo pairs in the
 Bolshoi simulations in that parameter range. Dark regions represent a high
 number of pairs. An absolute scaling for this shading can be obtained from
 the plots presented in the Right Column. The stars mark the location of the three
 LG pairs, each one coming from one of the constrained simulations. 
\emph{Right column} Integrated probability of these three different MAH
 times. The continuous black lines represent the results for the {\it
   Pairs} sample in the CLUES simulations. The {\it Isolated Pairs} sample from CLUES
 is represented by the thick dashed lines. The results from the {\it
   Isolated Pairs} samples in eight
 sub-volumes of the Bolshoi simulation are represented by the thin continuous
 grey lines. The thick continuous lines represent the results for the
 {\it LG} sample.
 The distributions from the {\it Pairs} and {\it Isolated Pairs}
 control samples are basically indistinguishable. In other words,
 detailed selection criteria for halo pairs, based on isolation only,
 do not narrow down significantly the range of dark matter halo
 assembly properties. 
\label{fig:surface}}
\end{figure*}

Four different samples of halos are constructed here, in a nested hierarchy in
which the first sample contains the second which contains the third. The
fourth sample is the one that includes the 3 LGs. These are to be used to
study how the various criteria employed in constructing the samples affect the
MAH of its members. The first three samples are constructed also from the
Bolshoi simulation, and are used to look for possible biases in the constrained
simulations.

The first sample we define consists of all halos in the mass range
$5\times 10^{11}$\hMsun $<M_{h}<$$ 5\times 10^{12}$ \hMsun
\citep{2010MNRAS.406..264W}. We refer to
this set as the {\it Individuals} halo sample. 

The second is a sample of halo pairs. Two halos, $H_{A}$ and $H_{B}$,
from the {\it Individuals} sample are considered a pair if and only if
halo $H_{B}$ is the closest halo to $H_{A}$ and
vice-versa. Furthermore, with respect to each halo in the pair there
cannot be any halo more massive than $5.0\times10^{12}$\hMsun closer than
its companion \citep{2004AJ....127.2031K}. We do not apply any further dynamical restrictions. For
instance an element in this sample may be a pair of halos that are
infalling into a cluster and are coincidentally close  to each
other. We refer to this set as the {\it Pairs} sample.    

The third is a sample of isolated pairs. We construct it by imposing
additional conditions on each member of the previous sample. These conditions
are defined to obtain a LG-like halo pair according to a series of
requirements that follow the lines of  \cite{1997NewA....2...91G},
\cite{2005MNRAS.359..941M}  \cite{2007MNRAS.378.1601M}. We will refer
to this sample as the {\it Isolated Pairs} sample. The conditions are
the following: 
z
\begin{enumerate}

\item[a)]The distance between the center of the halos is smaller than $0.7$
  \hMpc \citep{2005ApJ...635L..37R}.
\item[b)]The relative radial velocity of the two halos is negative.
\item[c)]There must not be objects more massive than either of the LG halos within 
a radius of $2$\hMpc from each object \citep{2009MNRAS.395.1915T}.
\item[d)]There must not be a halo of mass $> 5.0\times 10^{13}$\hMsun within a
  radius of $5$\hMpc with respect to each halo center \citep{2004AJ....127.2031K}.

\end{enumerate}

The  final fourth sample contains the three objects that fulfil the criteria
of third sample, and are located at about $10 \hMpc$ 'south' of the Virgo
cluster in the Supergalactic Plane. This  sample is referred to as {\it LG}.

We build these three samples both from the CLUES and Bolshoi simulations.
A short summary description of each sample is contained in Table \ref{table:samples}.

\section{Results}

\label{sec:results}
The backbone of our analysis is the study of the MAH of halos in the
mass range $\big(5.0\times 10^{11}\ - \  5.0\times
10^{12}\big)$\hMsun. Our results must be described in the 6
dimensional parameter space, spanned by  the three characteristic
times of the two halos, dubbed as MW and M31. The distribution of
$\tau_M$, $\tau_F$ and $\tau_A$ of the three different samples is
studied in \S\ref{sec:assembly} and the possible dependence of these
distributions on the ambient density around the LGs and the mass ratio
of the MW and M31 members of the LGs in \S\ref{sec:density}.   

\subsection{Mass accretion history of the different samples}
\label{sec:assembly}

Figure \ref{fig:integrated_A_B} presents distribution of $\tau_M$,
$\tau_F$ and $\tau_A$ for  the {\it Individuals} and {\it Pairs}
samples  of both the CLUES and Bolshoi simulations. The distribution
with respect to the MW and M31 are virtually indistinguishable, and
the curves present both halos.  We calibrate the effect of cosmic
variance with the $(100\hMpc)^3$ volumes extracted from the Bolshoi
simulation. The results are overplotted as thin
magenta lines. The distribution of $\tau_M$ and $\tau_F$ are well within
the scatter of the sub-volumes, while the $\tau_A$ is somewhat out of
the range.   

We conclude that with  respect to the MAH, the constrained simulations  are
essentially unbiased with respect to the unconstrained one.   The interesting
fact that emerges here is the halos in the {\it Individuals}  and {\it Pairs}
sample share the same MAH, as expressed by the three times described here.

Figure \ref{fig:surface}  presents the main results of the paper. It
shows the distribution of the three  times for the different sample of
pairs of halos. The left column is made of three grey scale maps
describing the number of objects in the {\it Pairs}
sample in the subspace of $\big(\tau{_X^{M31}}, \tau{_X^{MW}}\big)$,
where $X=M, F, A$.   The shades represent the number of pairs
around a given region of parameter space calculated from the {\it
  Pairs} samples in the Bolshoi simulation.  The three different {\it
  LG} pairs are overplotted as stars.

The right hand side column of Figure \ref{fig:surface}  shows
the integrated  relative distribution of the halos in the three
different times of the {\it Pairs}, {\it Isolated Pairs} and {\it LGs}
samples. For the {\it LGs} this is further separated for  the MW and
M31 halos. The distribution of the {\it Isolated Pairs} of the Bolshoi
sub-volume is presented as well.

Two important comments can be made based on Figure \ref{fig:surface}. First,
we see that the times in the {\it LGs} sample are confined to a narrow range
compared to the broad {\it Pairs} sample. The merger, formation and assembly times
in this sample are confined within the range $9.5-12$ Gyr. Second, from the integrated
distribution, we infer that the {\it Pairs} and {\it Isolated   Pairs}
samples are virtually indistinguishable. This implies that the commonly used isolation criteria
\citep{1997NewA....2...91G,2005MNRAS.359..941M,2007MNRAS.378.1601M} do
not automatically produce the narrow parameter space occupied by the
{\it LG} pairs.

\subsection{The influence of the Local Matter Density and the Mass Ratio}
\label{sec:density}

The {\it Pairs} and {\it Isolated Pairs} samples are selected based on
isolation and dynamics. The similarity of  the distribution of the different
MAH times of the the different samples motivates us to look for the possible
dependence of these distributions on some other characteristics of the three
LGs.   In particular, the three LGs are found to share the two following
properties: the  mass ratio between the two halos and the matter over-density
in a sphere of $5$\hMpc radius \footnote{$\delta_{5}$   has been calculated from the total   mass in halos more massive than
  $1\times 10^{10}\hMsun$ contained within a   sphere of radius $5$\hMpc
  centered at the position of each halo.}, noted as $\delta_{5}$.   The values for the halo masses in the pairs and the local over-densities are listed in Table
\ref{table:assembly} together with the assembly, formation and last major merger
times.  
A series of sub-samples of the halos sample are constructed by requiring that the masses and mass ratios between the pairs are bounded by the LG limits or the values $\delta_5$.  These sub-sampling do not bias the LG-like objects towards the region of parameter space defined by the LG sample.

\section{Discussion}
\label{sec:discussion}

Three basic facts emerge from the results presented in the
previous section : a) the three LGs  share a common formation
history, b) this formation history is quiet out to at $\approx
(10-12)$Gyr and c) none of the selection rules applied here to the
pairs of halos have defined a sample of objects with MAH similar to
that of the three LGs. In what follows, we discuss the possible origin and the
predictable consequences of these facts.

\begin{table*}

\caption{\label{table:probability}Fraction of halos/pairs the different samples with
  times $\tau_{M}$, $\tau_{F}$ and $\tau_{A}$ located in the
  parameter space defined by the minima characteristic times of the LG
  halos in the constrained simulations. These minima  from the
  LGs are defined for each $\tau_{X}$ in two different ways: 1) as the mean value minus
  two times the standard deviation (see Table \ref{table:assembly})
  and 2) as the minimum value of all realisations. These minima times
  are denoted $\tau_{X}^{\prime}$ and $\tau_{X}^{\prime\prime}$
  respectively and are presented in the first rows. In the following
  rows, the first column describes the name and origin of
  the sample. The three following columns indicate the fraction of the
  total population with a $\tau_{X}$ larger than the calculated
  $\tau_{X}^{\prime}$ or $\tau_{X}^{\prime\prime}$ (in parenthesis).
  In the case of pairs
  samples, we require the times for both halos to be above the threshold. The last column refers to
  the three different $\tau_{X}$ being \emph{simultaneously} larger than the
  corresponding $\tau_{X}^{\prime}$ ($\tau_{X}^{\prime\prime}$).}
\begin{tabular}{lcccc}\hline
"Two sigma" bound & $\tau_{M}^{\prime}$ [Gyr]& $\tau_{F}^{\prime}$ [Gyr]& $\tau_{A}^{\prime}$ [Gyr]& \\
 & 9.3 & 9.0 &9.6 & \\\hline
"Minima" bound & $\tau_{M}^{\prime\prime}$ [Gyr]& $\tau_{F}^{\prime\prime}$ [Gyr]& $\tau_{A}^{\prime\prime}$ [Gyr]& \\
 & 9.8 & 9.3 &9.7 & \\\hline
Sample & $\tau_{M}\geq\tau_{M}^{\prime}$ ($\tau_{M}^{\prime\prime}$)& $\tau_{F}\geq\tau_{F}^{\prime}$ ($\tau_{F}^{\prime\prime}$)& $\tau_{A}\geq\tau_{A}^{\prime}$ ($\tau_{A}^{\prime\prime}$)& $\tau_{M,F,A}\geq\tau_{M,F,A}^{\prime}$ ($\tau_{M,F,A}^{\prime\prime}$)\\\hline
CLUES {\it Individuals}  & 0.24 (0.18)& 0.29 (0.24)& 0.85 (0.85)& 0.17
(0.12)\\
CLUES {\it Pairs} & 0.06 (0.03)& 0.09 (0.06)& 0.74 (0.74)& 0.03 (0.01)\\
CLUES {\it Isolated Pairs} & 0.06 (0.03)& 0.08 (0.05)& 0.70 (0.70)&
0.05 (0.03)\\
Bolshoi {\it Individuals}  & 0.23 (0.19)& 0.23 (0.23)& 0.87 (0.87)&
0.17 (0.12)\\
Bolshoi {\it Pairs}  & 0.05 (0.04)& 0.10 (0.05)& 0.76 (0.76)& 0.03 (0.02)\\
Bolshoi {\it Isolated Pairs}  & 0.05 (0.03)& 0.10 (0.06)& 0.73 (0.73)& 0.03 (0.01)\\\hline
\end{tabular}
\end{table*}

\subsection{On the Common Formation History}

Naively, one might hypothesise  that the fact that all three CLUES  LGs have a
common MAH, as defined here, is consistent with being drawn at random from the
sample of pairs,  i.e. the range of properties spanned by three random halo pairs can be
naturally expected to be narrow. This is the null hypothesis we test now.

What is the probability that 3 randomly selected pairs have MAHs
within the range of properties found for the LG? We compute this
probability based on the fraction of halos in the pair samples that
share the {\it LG} formation properties. 

We define first the minimal subspace that contains the 3 simulated LGs
by providing lower bounds on the different times describing the MAHs.
Table \ref{table:probability} lists the minimal last major merger,
formation and assembly look-back  times, where two options are taken to
estimate the minima. The first defines the "two sigma" bound, namely
the average value minus twice the standard deviation  of each time of
the 6 halos of the 3 LGs, the second takes the minimum value for each
time. 

The table provides the fraction of halos in the {\it Individuals}
sample satisfying each one of the conditions $\tau_X \geq
\tau_{X}^{\rm bound}$ independently and all of them simultaneously, where
$X=M$, $F$ and $A$ and the subscript ${bound}$ denotes the minimal bound of such
time.  We find that the fraction of {\it Individuals} in the quiet MAH subspace
is $f_{i}=0.17 (0.12)$ both in CLUES and Bolshoi for the first
(second) minima option.  If we consider now the halos either in the
{\it Pairs} or {\it Isolated Pairs} samples, only a fraction of $f_{p}=
0.03 (0.01)$ pairs are composed of halos that are both within the {\it
  LG} parameter space.  To a good approximation, the pair fraction can
be calculated as the individual fraction squared, $f_{p}\approx
f_{i}\times f_{i}$. This is the expected result under the assumption
that the assembly of the MW and M31  are independent.

The probability of randomly selecting three random halo pairs and
having them within the range of parameters defined by the {\it LG} can
be calculated as $p_{LG}=f_{p}^3\approx 2.7\times 10^{-5} (1.0\times 10^{-6})$.  This small
probability is a consequence of having found 3 halo pairs within a set
of properties shared by $0.17 (0.12)$ of the  total population of
halos. If we consider pairs with a range of   desired properties
within shared by, say, $0.68$ of the halos in the total population (the
fraction within one standard deviation around the mean), the
probability of finding three pairs inside that range  would be
$p_{1-\sigma}(0.68 \times 0.68)^{3}\approx 0.1$. 

Comparing the results of the probabilities  $p_{LG}$ and
$p_{1-\sigma}$, the null hypothesis can be safely rejected. It is
highly unlikely that the three randomly selected  pairs show a
narrow range of properties as in the case of the {\it LG} sample. 

Both the {\it ab-initio} and {\it   post-factum} constraints imposed
on the  LG yield a {\it LG} sample with very similar MAHs. In the CLUES
simulations only the large and mid-scales are effectively constrained
by the data leaving the galactic and smaller scales effectively
random. It follows that the MAH of objects similar to the LG is
strongly affected by their environment. To what extent this is valid
for DM halos in general remains an open question.

\subsection{On the Quietness of the Formation History}

We established in the previous sections that the MAHs are quiet out to $\approx
(10-12)$Gyr, and that none of the selection rules applied here to the pairs
of halos have defined a sample of objects with MAH similar to that of
the three LGs.  

The last point is consistent with the results previous studies that have
approached the same question of estimating a possible bias of the LG with
respect to a general halo population
\citep{2009MNRAS.395..210D,2010MNRAS.406..896B}. These studies apply
isolation criteria on scales of $1$\hMpc over halos in the mass range
we study here, and find as well that no significant bias is introduced
in the isolated halo population with respect to the parent halo
population.  

The parameter subspace defined  by the three {\it   LG}
cannot by explained either in terms of the isolation criteria listed
at the end of \S\ref{sec:sample} or by adding constraints on
the values of the local over-density on $5$\hMpc scales. The
properties of the dynamical environment, common to all the CLUES
simulations and provide the quiet formation history for a LG, remain
to be found. Ideally, that result should be expressed in a suitable
form to search for LG pairs in an unconstrained simulation.

Is the observed Local Group biased in the same manner? We cannot
provide the answer to that question with the simulations we
present in this paper. Nonetheless,  the theoretical predictions
we show here for the dark matter assembly in the LG seem to be in
agreement with the disk dominated morphology of MW and M31.

\subsection{The Connection with the Observed Local Group}
\label{sec:observations}

The most distinct feature of the MW and M31 is that both galaxies have a
disk dominated morphology. It is often mentioned that abundant mergers, which are
presumed to destroy the disk and be source of morphological change, are expected on all
mass scales in the hierarchical picture of galaxy formation of \LCDM\;
generating a possible contradiction with the abundance of disk galaxies
in the local Universe and, in particular, with the fact that the MW
and M31 are disk galaxies
\citep{1992ApJ...389....5T,1993ApJ...403...74Q,2008ApJ...688..254K}.

Our results provide new theoretical evidence that the MW and M31
could be expected to be disk dominated galaxies in \LCDM . From the
results presented here, we have found that the last merger started on
average $11$ Gyr ago.  At these redshifts the mass of the MW host halo  is
$1-4\times 10^{11}$\hMsun, its virial velocity is $\approx 200$ km/s
and its virial radius $\approx 0.1$ \hMpc. Using these quantities and
Eq. \ref{eq:infall} we estimate the final infall time for the
satellite to be $\approx 3.5$ Gyr, reaching the center $\approx 7.5$
Gyr ago. This quiet history should favour the survival of a disk
formed in the halo \citep{2011arXiv1103.6030G}. Although, detailed estimations on these matters
might have to include the inflow of gas into the disk
\citep{2009MNRAS.396..696S}.  

A distinct and well characterised feature of the MW is the thick
disk. This disk component of the MW has been known for more than 25
years \citep{1983MNRAS.202.1025G}. The thick disk contains  a  
population of stars with different kinematics, spatial distribution,
ages and chemical enrichment compared to the thin galactic disk.  Although M31
seems to have a similar component \citep{2011MNRAS.tmp..248C}, the
observational and theoretical work on the MW's thick disk has a long history, and
its origin can therefore be discussed in greater detail.
One of the possible formation scenarios for the MW  thick disk is an  in-situ formation
during/after a gas rich merger \citep{2009MNRAS.400L..61S}. The analysis of
the  orbital eccentricity of stars based on  RAVE and SDSS data supports
the gas-rich merger mechanism
\citep{2010ApJ...725L.186D,2011MNRAS.tmp..260W}. In our results the last
merger reaches the center $\approx 7.5$ Gyr ago, close to the look-back time of
$8$Gyr as required by the in-situ formation scenario.

\section{Conclusions}
\label{sec:conclusions}

We use constrained simulations of the local Universe to study the
dark matter mass aggregation history (MAH) of the Local Group (LG). Two basic
questions motivate this study: 1. To what extent the simulated LGs can
account for the observed structure of the MW and M31 galaxies? Namely,
if the disk dominated morphology implies that the MW and M31 halos had
a quiet MAH over the last $\approx 11$Gyrs, can simulations
recover this recent quiet history? 2. Does this quiet MAH arise from
the intrinsic properties of the DM halos, or is it induced by
environment within which the LG is embedded? Is the implied MAH of
the LG triggered by the large and meso-scales, or is it induced by the
small, i.e. galactic and sub-galactic, scales?

The methodology adopted here is to use constrained simulations of the
local Universe, designed to reproduce the large and meso-scales of
the LG environment, and search for halos that resemble the actual
LG. The identification of a pair of halos as a LG-like object is based
on a set of isolation and dynamical criteria, all formulated by their
redshift zero structure, in complete ignorance of their formation history. A
LG-like object that is found close to the actual position of the
observed LG with respect to the large scale structure environment is defined here as a LG. By construction a constrained
simulation can have only one LG or none at all. Indeed, out of a suit
of 200  constrained simulations only 3 harbour a LG. Controlled samples
of individual halos and pairs have been constructed as reference
samples. The analysis has been extended to the unconstrained Bolshoi
simulation that is used here for an unbiased reference
\citep{2010arXiv1002.3660K}.

The construction of the identification of the 3 LGs is done
independently of the MAH of the halos. Yet, the MW's and M31's halos
of the 3 LGs all have a common quiet MAH, defined as having the last
major merger, formation and assembly look-back time extending over
$\approx (10\ -12)$ Gyr. This  quiet formation history of the simulated
LGs can help to explain the disk dominated morphology of the MW and
M31, adding evidence to the internal instability origin of the
spheroidal component of the MW \citep{2010ApJ...720L..72S}.  Based on
measurements of the eccentricity of orbits in the MW, it has been
recently claimed
\citep{2009MNRAS.400L..61S,2010ApJ...725L.186D,2011MNRAS.tmp..260W},
that a  rich merger taking place 10.5 to 8 Gyr ago is a favoured
mechanism  explain the thick disk in the MW  \cite{2004ApJ...612..894B}.  Our
finding of a quiet MAH of the LG provides a suitable platform for such
a process to take place.

The LG halos are assumed here to be selected from FOF halos in the
mass range $5\times 10^{11}\hMsun <M_{h} < 5\times 10^{12}\hMsun$ at
$z=0$. Between $12\%$ and $17\%$ of these halos are found 
to have a quiet MAH, depending on the detailed definition of the quiet
parameter space. From this point of view the MW
and M31 halos are not  rare. However, how likely is a pair of halos to
have such a quiet history, shared by both halos? Making the naive
null assumption that the MAH of a halo is an intrinsic property of a
halo independent of its environment then the fraction of pairs should be the
product of the fractions for a single halo. Indeed, the {\it Pairs}
sample drawn out of the Bolshoi simulation confirms this assertion,
finding that between $1\%$ to $3\%$ of the pairs have as quiet an MAH as
the LG systems do. The probability of having selecting 3 pairs randomly
and finding them with a quiet MAH is on the order of $\sim 10^{-5}$.

Next, we look for what dynamical or environmental property determines
rethe MAH of a LG-like object.  We find here  that the mere pairing of
the MW-like halos does not affect the MAH fiducial times. Imposing 
the isolation and dynamical constraints that define the {\it
  Isolated Pairs} sample  does not affect it either. This leaves us
with an open question as to what determines the MAH of halo pairs
similar to the LG. The one hint that we have is that all of the three LGs
reside in the same large and meso-scale environment. We speculate that
the cosmic web plays a major role in shaping the MAH of LG-like
objects,  although it is not yet clear what mechanism is
responsible. A larger sample of constrained LGs is needed to confirm and
further explore the reasons behind this result. 

\section*{Acknowledgements}
We acknowledge stimulating discussions with Cecilia Scannapieco and Noam
I. Libeskind. Y.H. has been partially supported by the ISF (13/08) and the
Johann Wempe Award by the AIP. We acknowledge the use of the CLUES data storage
system EREBOS at AIP. GY would like to thank the MICINN (Spain) for  financial
support under project numbers FPA 2009-08958  AYA 2009-13875-C03 and the SyeC
Consolider project CSD 2007-0050. The simulations were performed at the Leibniz
Rechenzentrum Munich (LRZ) and at  Barcelona Supercomputing Center (BSC). We
thank DEISA for giving us access to computing resources in these  through the
DECI projects SIMU-LU and SIMUGAL-LU.

\bibliographystyle{mn2e}

\end{document}